\documentclass{PoS}

\title{More about the new non-selfdual \\
       axially symmetric $SU(2)$ calorons}

\ShortTitle{Non-selfdual calorons}

\author{\speaker{Ya.~M.~Shnir}\\
        Department of Computer Sciences, National University of Ireland, Maynooth, Co. Kildare, Ireland,  \\
	Dublin Institute of Advanced Studies, School of Theoretical Physics, 10 Burlington Road, Dublin, Ireland \\
        E-mail: \email{shnir@maths.tcd.ie}}

\author{E.-M.~Ilgenfritz\\
        Institut f\"ur Physik, Humboldt-Universit\"at zu Berlin, Newton-Str. 15, D-12489 Berlin, Germany,  \\
	Institut f\"ur Physik, Karl-Franzens-Universit\"at Graz, Universit\"atsplatz 5, A-8010 Graz, Austria \\
        E-mail: \email{ilgenfri@physik.hu-berlin.de}}

\abstract{We describe the recently found non-selfdual axially symmetric 
caloron solutions of $SU(2)$ gluodynamics with trivial holonomy.
We present the local Polyakov loop together with the action and topological 
charge density. Different from the well-known KvBLL calorons, the 
calorons considered here are composed out of pseudoparticle 
constituents carrying {\it integer} topological charge. 
The loci of Polyakov loop $L = -1$ correspond to the loci of Higgs field
zeroes for corresponding solutions of the Yang-Mills Higgs system. 
For certain parameters pointlike monopole pairs turn into rings. } 

\FullConference{8th Conference Quark Confinement and the Hadron Spectrum \\
	 September 1-6 2008\\
	 Mainz, Germany}

\begin{document}

\section{Introduction}

The relations between the properties of selfdual BPS monopole
solutions~\cite{Bogomolny:1975de} and instantons~\cite{Instanton} 
attracted a lot of interest over the last decade. It was shown that exact 
caloron solutions, {\it i.e.} time-periodic instantons at finite 
temperature on $\mathbb{R}^3\times S^1$, for which the temporal gauge 
field component $A_0$ approaches some constant at spatial 
infinity~\cite{HarrShep78,Gross83}, 
$A_0 \to 2 \pi i \omega = 2 \pi i  \omega^a \sigma^a$, are composed out 
of Bogomol'nyi-Prasad-Sommerfeld (BPS) monopole 
(and antimonopole) constituents~\cite{Baal}. This time-periodic 
instanton eventually corresponds to a non-trivial Polyakov loop 
(holonomy around $S^1$) at spatial infinity. In the periodic gauge 
$A_\mu(\mathbf{r}, x_0+T)= A_\mu(\mathbf{r}, x_0)$ the Polyakov loop 
is defined as
\begin{equation}
\label{eq:Polakov-loop}
{\cal P}({\bf r}) = \lim_{r\to\infty} P \exp\left(
\int\limits_{0}^{T} A_0({\bf r},x_0)d x_0\right)\, , 
\end{equation} 
where $T$ is the period in the imaginary time direction, which is related
to the temperature $\Theta$ through $T=1/\Theta$. 
The symbol $P$ denotes path ordering. 

The property of selfduality allows to apply the very powerful
ADHM-Nahm formalism~\cite{ADHM} to obtain various exact multi-caloron 
configurations~\cite{Baal,BaalFalk} and to analyse the properties of 
the BPS monopole constituents. 
In particular, it was shown that, if the size of a charge one $SU(2)$
caloron is getting larger than the period $T$, the caloron is splitting 
into constituents, {\it i.e.} 
the monopole-antimonopole pair becomes visible through well-separated
lumps of action. The properties of the saddle point solutions
in the related $SU(2)$ Yang-Mills-Higgs (YMH) model were discussed
first by Taubes~\cite{Taubes81}, and various monopole-antimonopole 
YMH systems were constructed numerically in Refs.~\cite{Rueber,mapKK,KKS}, 
both in the BPS limit and beyond.

However, besides the selfdual instantons, also solutions of the
second order Euler-Lagrange equations of the Euclidean Yang-Mills
(YM) theory are known~\cite{non-sd-Instanton}. Thus, a non-selfdual 
instanton-antiinstanton pair static configuration has been 
constructed~\cite{TR06}, which represents a saddle point configuration, 
{\it i.e.} a deformation of the topologically trivial field.

Recently new static and axially symmetric $SU(2)$ YM caloron solutions 
on $\mathbb{R}^3\times S^1$ with trivial holonomy were constructed 
numerically~\cite{S07}. These regular field configurations are labeled 
by two integers $n$ and $m$, analogously to their counterparts in the 
YMH system, the monopole-antimonopole chains and the circular 
vortices~\cite{KKS}. Similar to the case of axially symmetric instantons 
discussed in~\cite{TR06}, only the $m=1$ solutions are selfdual. 
The calorons labeled by $m \ge 2$ are non-selfdual. The latter are also 
composed of constituents and correspond to the monopole-antimonopole 
chains and/or to the vortex-like solutions of Ref.~\cite{S07}.

In this talk we briefly describe the properties of the new non-selfdual,
axially symmetric caloron solutions of the second order field equations, 
adding here the results of the numerical evaluation of the holonomy in 
the caloron background to the main characteristics.

\section{The Euclidean $SU(2)$ action and the axially symmetric ansatz}

We consider the usual $SU(2)$ YM action
\begin{equation} 
\label{eq:S}
S = \frac{1}{2}  \int d^4 x \rm{Tr} \left( F_{\mu\nu} F_{\mu\nu}\right)
= \frac{1}{4}  \int d^4 x \left(  F_{\mu\nu} \pm {\widetilde  F}_{\mu\nu}\right)^2
\mp \frac{1}{2}  \int d^4 x \rm{Tr} \left(  F_{\mu\nu}{\widetilde  F}_{\mu\nu}\right)
\, .
\end{equation}
We work in Euclidean space $R^3 \times S^1$ with compactified time 
running in $x_0 \in [0,T]$. The gauge coupling is put equal to $e^2=1$. 
The $su(2)$ gauge potential is $A_\mu = A_\mu^a \tau^a/2$, and the field 
strength tensor is
$F_{\mu\nu} = \partial_\mu A_\nu - \partial_\nu A_\mu + i[A_\mu, A_\nu]$.
The topological charge 
$Q= \frac{1}{32\pi^2} \varepsilon_{\mu\nu\rho\sigma} \int d^4x
\rm{Tr}  F_{\mu\nu} F_{\rho\sigma}$. is defined by the integral over all
space-time. Only for selfdual configurations the bound for the action
$S \ge 8 \pi^2 |Q|$ (and a similar local bound for the densities) is saturated.

To construct the new regular caloron solutions of the corresponding
\emph{second order} field equations and in order to investigate the 
dependence of these solutions on the boundary conditions, we employ 
the axially symmetric ansatz for the gauge field
\begin{eqnarray} 
\label{eq:ansatz}
A_\mu dx^\mu =
\left( \frac{K_1}{r} dr + (1-K_2)d\theta\right)\frac{\tau_\varphi^{(n)}}{2e}
&-&n \sin\theta \left( K_3\frac{\tau_r^{(n,m)}}{2e}
                     +(1-K_4)\frac{\tau_\theta^{(n,m)}}{2e}
\right) d\varphi \, ; \nonumber \\
A_0 = A_0^a\frac{\tau^a}{2} & = &
\left(K_5\frac{\tau_r^{(n,m)}}{2}+ K_6\frac{\tau_\theta^{(n,m)}}{2} \right) \,  , 
\end{eqnarray}
that was previously applied to the YMH system~\cite{KKS}. 
The ansatz is written in the basis of $su(2)$ matrices
$\tau_r^{(n,m)},\tau_\theta^{(n,m)} $ and $\tau_\varphi^{(n)}$ which
are defined as the dot product between the Cartesian vector of Pauli
matrices ${\vec \tau}$ and the spatial unit vector that
generalizes the local dreibein ${\vec x}_i/|{\vec x}_i|=\hat{e}^{(1,1)}_i$
of tangential vectors ($i=r,\theta,\phi$ denote the tangential directions).
The generalization proceeds analogously to the radial one, 
$\hat{e}^{(1,1)}_r \to \hat{e}^{(n,m)}_r = \left(\sin(m\theta) \cos(n\phi),\sin(m\theta) \sin(n\phi), 
\cos(m\theta)\right)$, by changing $\phi \to n\phi$ and $\theta \to m\theta$.
The gauge field functions $K_i(r,\theta)$ ($i=1,\dots,6$) depend only 
on the spherical coordinates $r$ and $\theta$ 

Substitution of the axially symmetric ansatz (\ref{eq:ansatz}) into the
definition of the topological charge $Q$ yields 
$Q = \frac{n}{2} \left[1-(-1)^m\right]\, , $
similar to~\cite{KKS,TR06}. Thus, the configurations labeled by
even $m$ belong to the topologically trivial sector and represent
saddle point solutions.

To satisfy the condition of finiteness of the total Euclidean
action (\ref{eq:S}), we additionally require that the field strength 
vanishes 
as $\rm{Tr} (F_{\mu\nu} F_{\mu\nu}) \to O(r^{-4})$ with $r \to \infty$. 
In the regular gauge the value of the $A_0$ component of the gauge
potential approaches a constant at spatial infinity, {\it i.e.} 
$A_0 \to \frac{i \beta}{2}\tau_r^{(n,m)}$. This corresponds to the 
holonomy at infinity (\ref{eq:Polakov-loop}), 
\begin{equation}
\label{eq:Polakov-loop-1} 
\frac{1}{2}~\rm{Tr}~ {\cal P}({\vec r}) \to 
\frac{1}{2}~\rm{Tr}~ \exp\left(\frac{i \beta T}{2}\tau_r^{(n,m)}\right) 
= \frac{1}{2}\rm{Tr}~ U \exp \left( \frac{i\beta
T}{2}\tau_z \right) U^{-1} = \cos \frac{\beta T}{2} \, , 
\end{equation}
where $U \in SU(2)$ and $\beta \in [0; 2\pi/T]$. 

We consider now deformations in the topologically trivial sector
and deformations of the caloron solution with trivial holonomy
at infinity~\cite{HarrShep78,Chakrabati87}. The latter is defined
as a time-periodic array of $Q=1$ or $Q=-1$ instantons, located 
along the Euclidean time axis (with distance $T$). A possible
generalization of this solution corresponds to a time-periodic 
array of instantons with general charge $Q$~\cite{Chakrabati87}. 
Now we will not require anymore that the gauge field should be 
selfdual such that, generically, 
$F_{\mu\nu}(x) \ne \pm \widetilde{F}_{\mu\nu}(x)$.

The regular caloron solutions with finite action density and proper 
asymptotic behavior can be constructed numerically by imposing 
boundary conditions~\cite{S07}
and solving the resulting system of six coupled 
non-linear partial differential equations of second order~\cite{S07}.
As usual, to obtain a regular solutions we have to 
satisfy the gauge 
condition $\partial_r A_r + \partial_\theta A_\theta = 0$~\cite{KKS}. 
We also introduce a compact radial coordinate 
$x=r/(1+r) \in [0:1]$. The numerical calculations were performed with 
the software package FIDISOL based on the Newton-Raphson iterative 
procedure. 

\section{Discussion of the solutions}

The simplest class of solutions corresponds to $m=1$. 
It turns out that, similar to~\cite{TR06}, these solutions are
selfdual. We check this conclusion by numerical calculation of
the integrated action density, as well as direct substitution of
the solutions into the first order equation expressing selfduality.
Furthermore, the $m=n=1$ solution is nothing but the 
Harrington-Shepard~\cite{HarrShep78}
spherically symmetric finite temperature solution  
of unit topological charge. The solutions with $m=1$ and $n \ge 2$ are 
of reduced, {\it i.e.} axial symmetry. 
Their distribution of action has the shape of a torus around the $z-$axis. 

%%%%%%%%%%%%%%%%%%%%%%%%%%%%%%%%%%%%%%%%%%%%%%%%%%%%%%%%%%%%%%%%%%%%%

\begin{figure}
\begin{center}
  \includegraphics[height=.30\textheight, angle =-90]{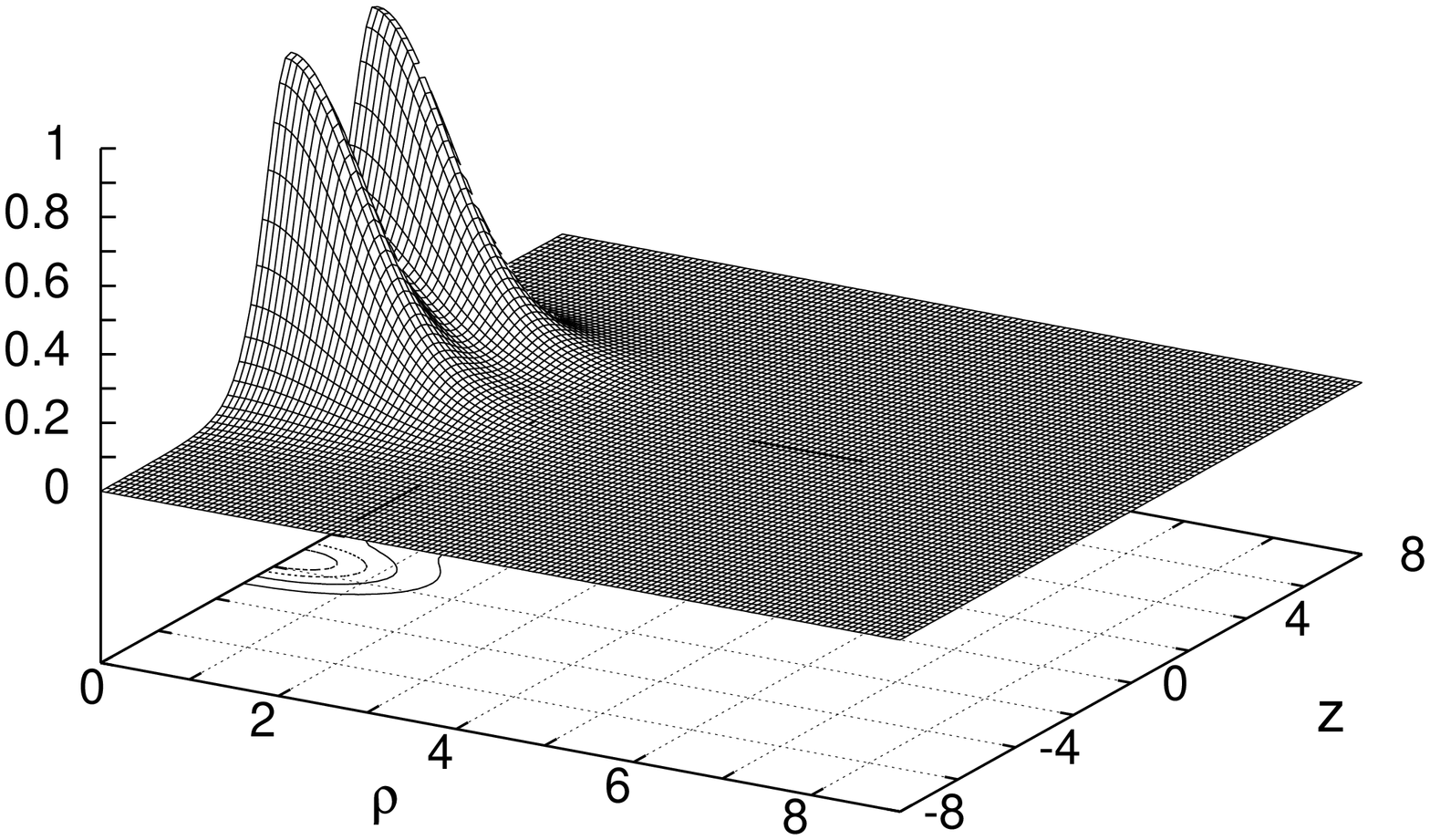}
  \includegraphics[height=.30\textheight, angle =-90]{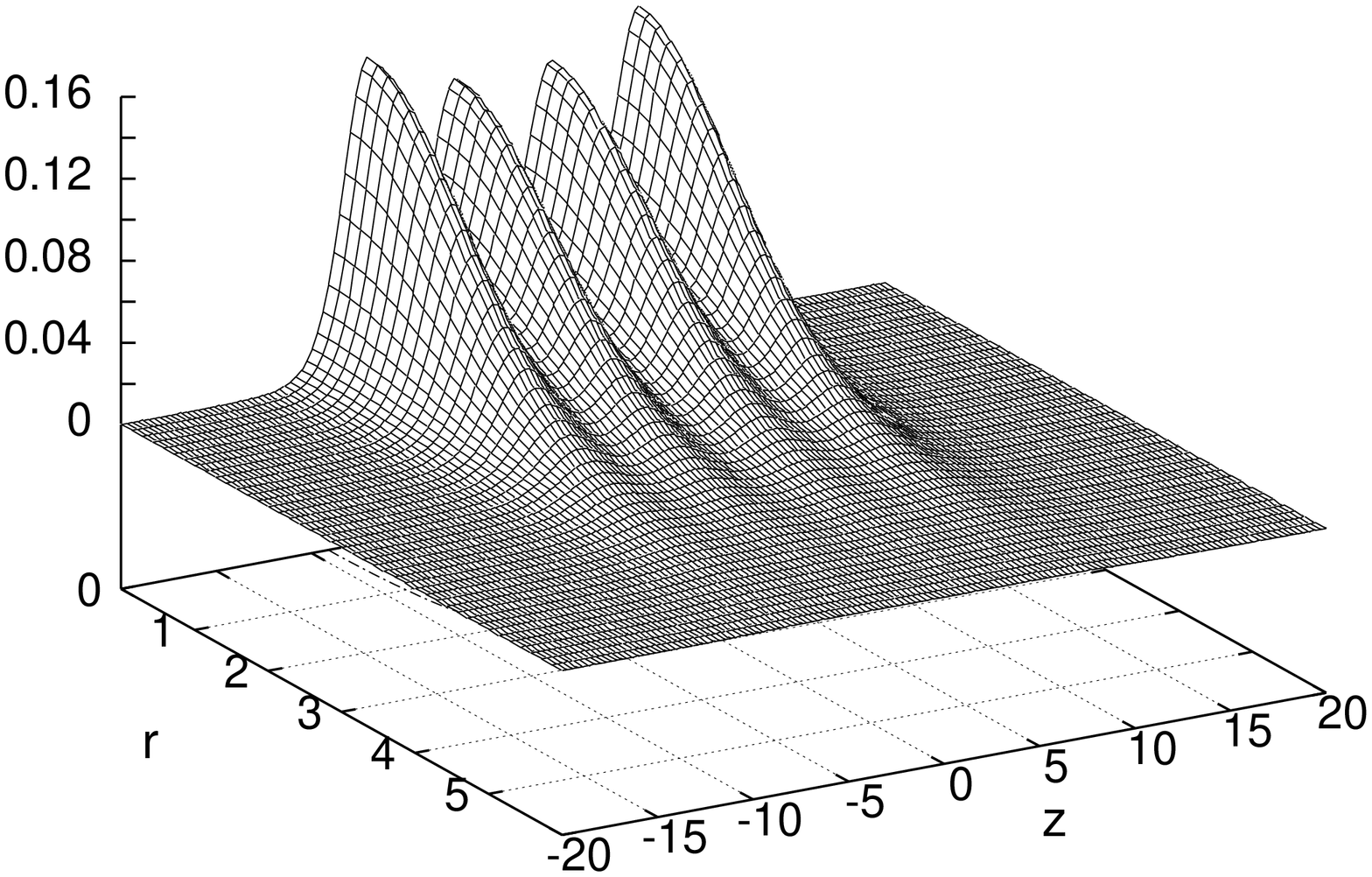}
  \includegraphics[height=.30\textheight, angle =-90]{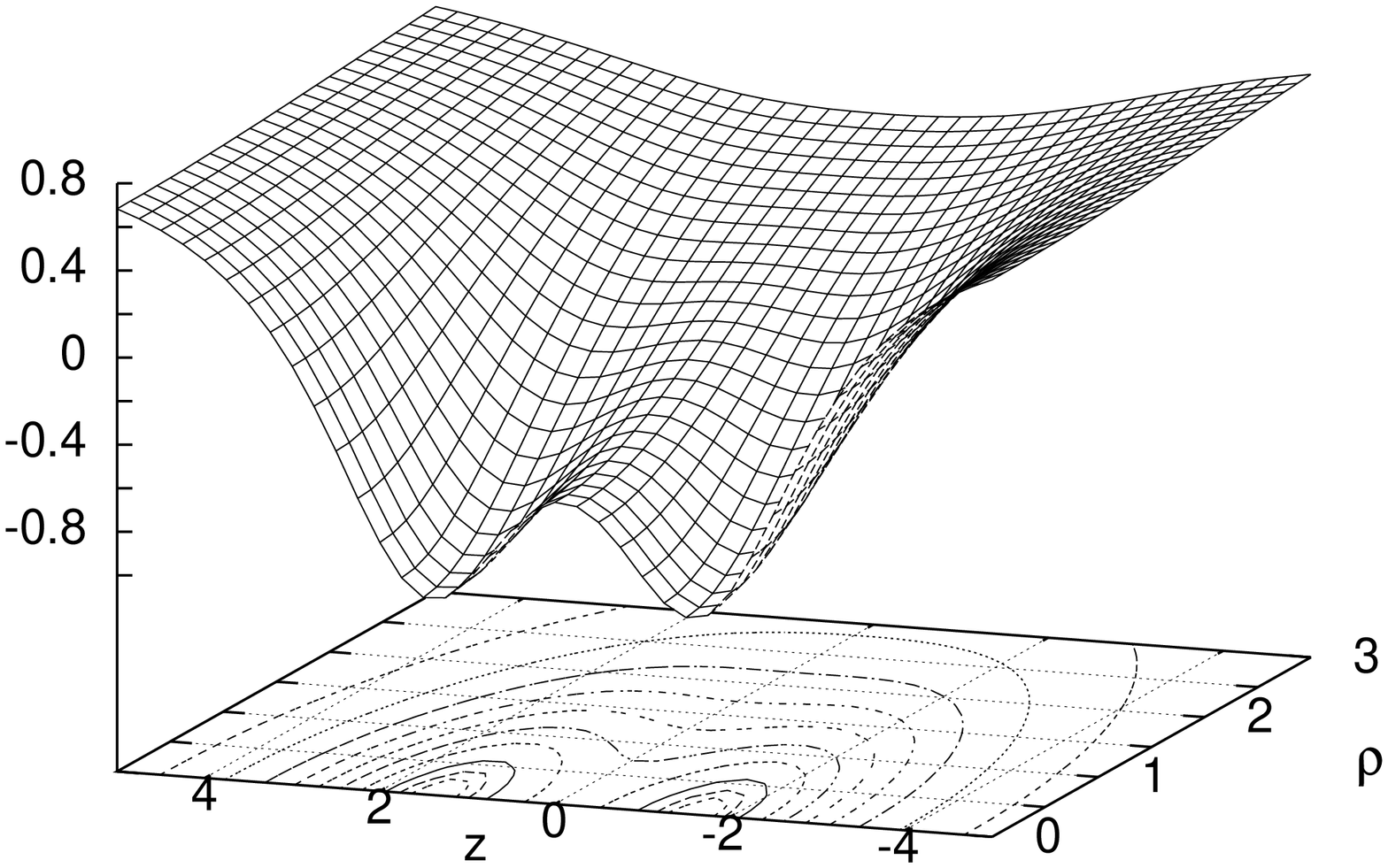}
  \includegraphics[height=.30\textheight, angle =-90]{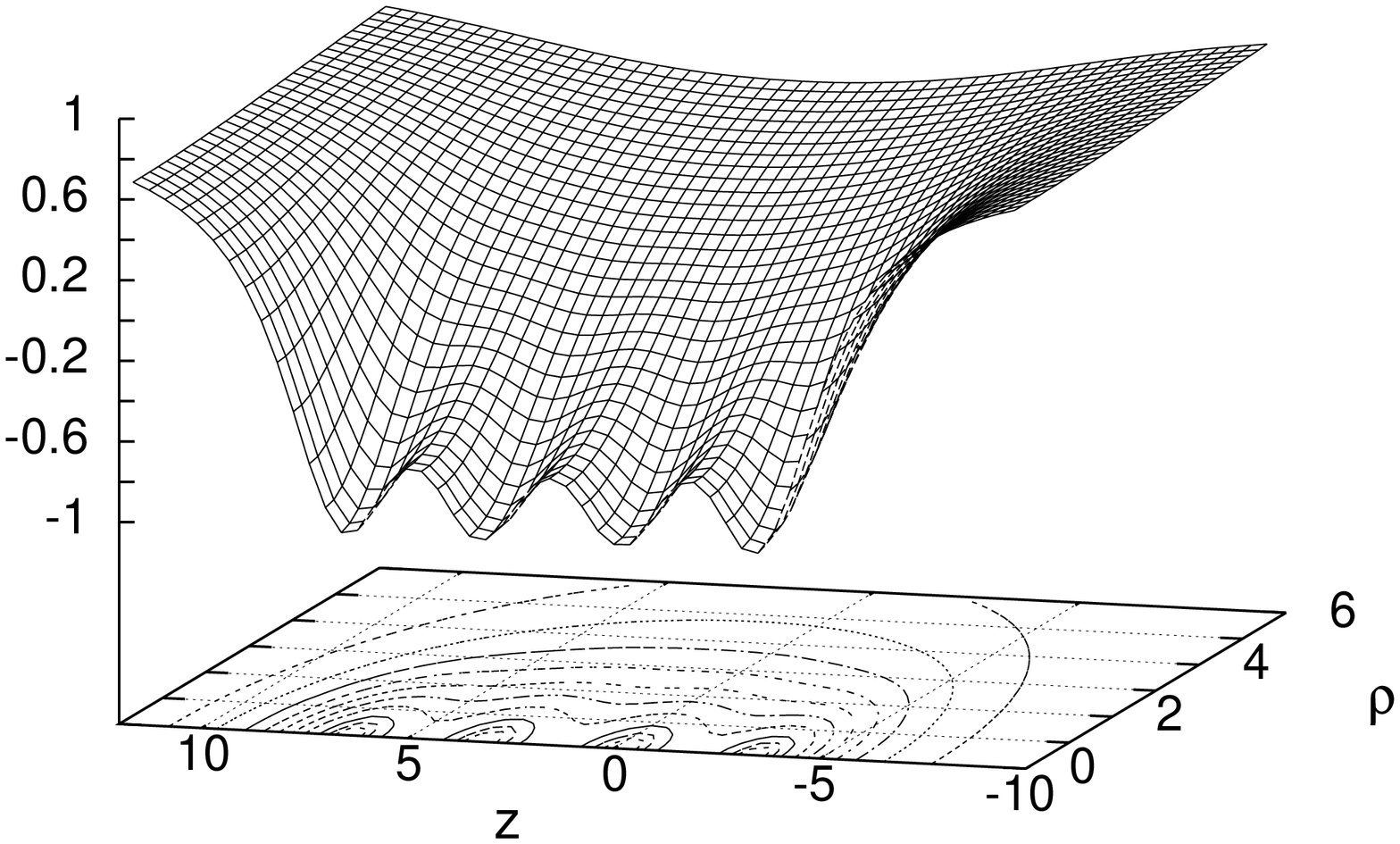}
\end{center}
\vspace{-0.2cm}
\caption{The action density for the $m=2$ (upper left) and $m=4$ 
(upper right) caloron chains, both with winding number $n=1$, is 
shown vs. cylindrical coordinates $z$ and $\rho$. 
The Polyakov loop is plotted below.}
\label{fig:figure1}
\end{figure}

The $m \ge 2$ configurations do not satisfy the first order (selfduality)
equations. Similarly to their counterparts 
in YMH theory~\cite{KKS}, the solutions with $n=1$ and $m=2,3,4 \dots$ 
represent time-periodic arrays of finite-length chains containing 
instantons and anti-instantons of $Q = \pm 1$ topological charge in 
alternating order, located along the spatial symmetry axis with $m$ 
clearly separated maxima of the action density (see Fig \ref{fig:figure1}). 
The topological charge density possesses $m$ local extrema along the 
$z$ axis, whose locations coincide with the maxima of the action density. 
Thus, one can distinguish $m$ individual constituents.

%%%%%%%%%%%%%%%%%%%%%%%%%%%%%%%%%%%%%%%%%%%%%%%%%%%%%%%%%%%%%%%%%%%%%
%%%%%%%%%%%%%%%%%%%%%%%%%%%%%%%%%%%%%%%%%%%%%%%%%%%%%%%%%%%%%%%%%%%%%

To compute the local Polyakov loop at a given point ${\vec x}$ 
we note that in the static regular gauge the temporal component 
$A_0$ of the gauge potential approaches a constant at spatial infinity 
although the holonomy is trivial. To check it, we make use of the 
numerical solutions found above and perform the integration in the 
exponent (in the static gauge). 
One gets $\rm{Tr}~ {\cal P}({\vec r}) = \cos ||A_0({\vec r})|| $ 
which agrees with (\ref{eq:Polakov-loop-1}) asymptotically. 

For the solutions of the chain type with $n=1$ and 
$m=2,3,4 \dots$ the Polyakov loop takes extremal values 
${\cal P}({\vec x_i})=-1$ (opposite to the asymptotic value 
${\cal P}(|{\vec x}| \to \infty)=1$) on the symmetry axis where the 
constituents are located at ${\vec x_i}=\left(0,0,z_i\right)$ (see 
Fig. \ref{fig:figure1}). 
In this sense the local Polyakov loop 
perfectly corresponds to the action density. 

\begin{figure}
\begin{center}
  \includegraphics[height=.30\textheight, angle =-90]{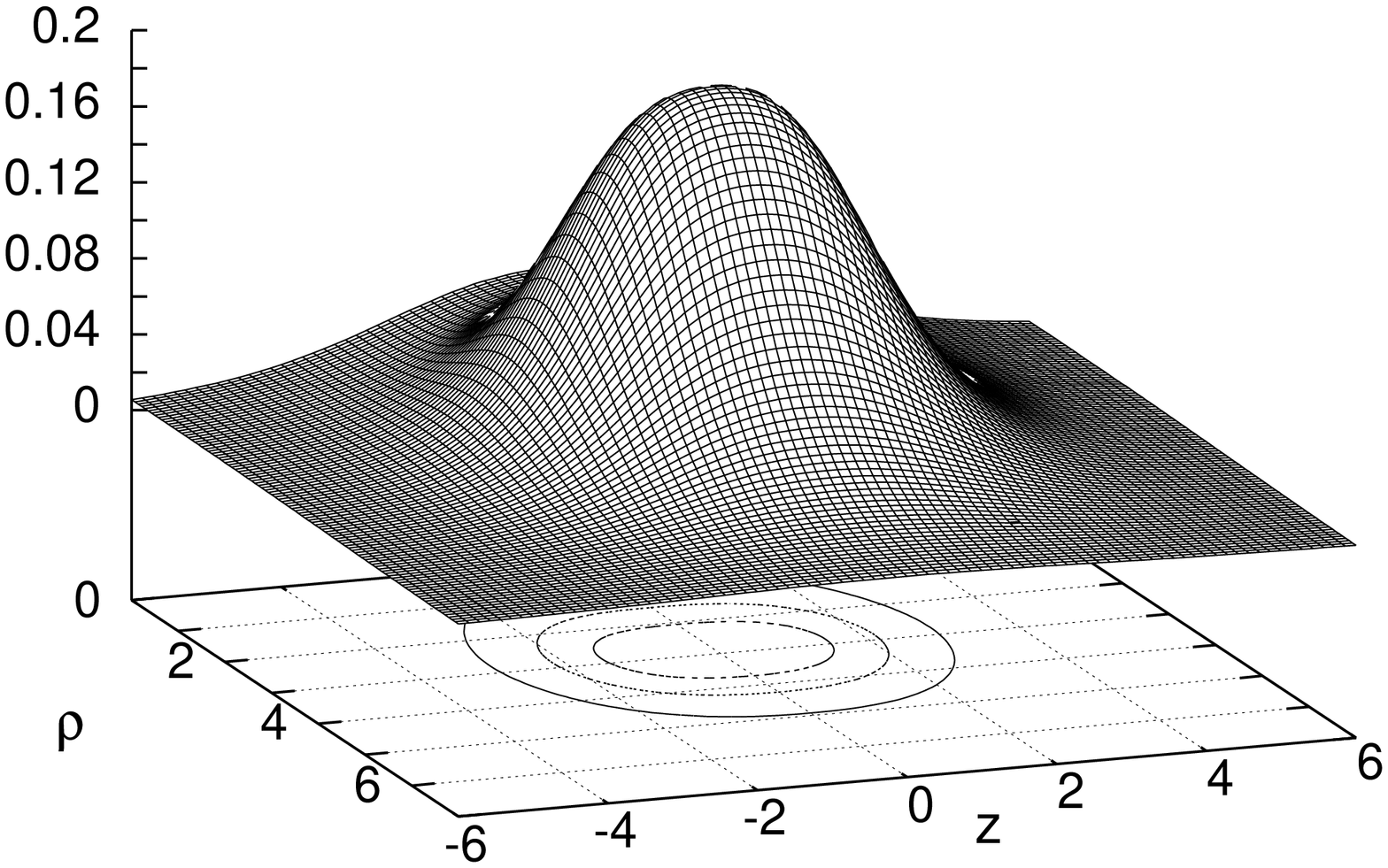}
  \includegraphics[height=.30\textheight, angle =-90]{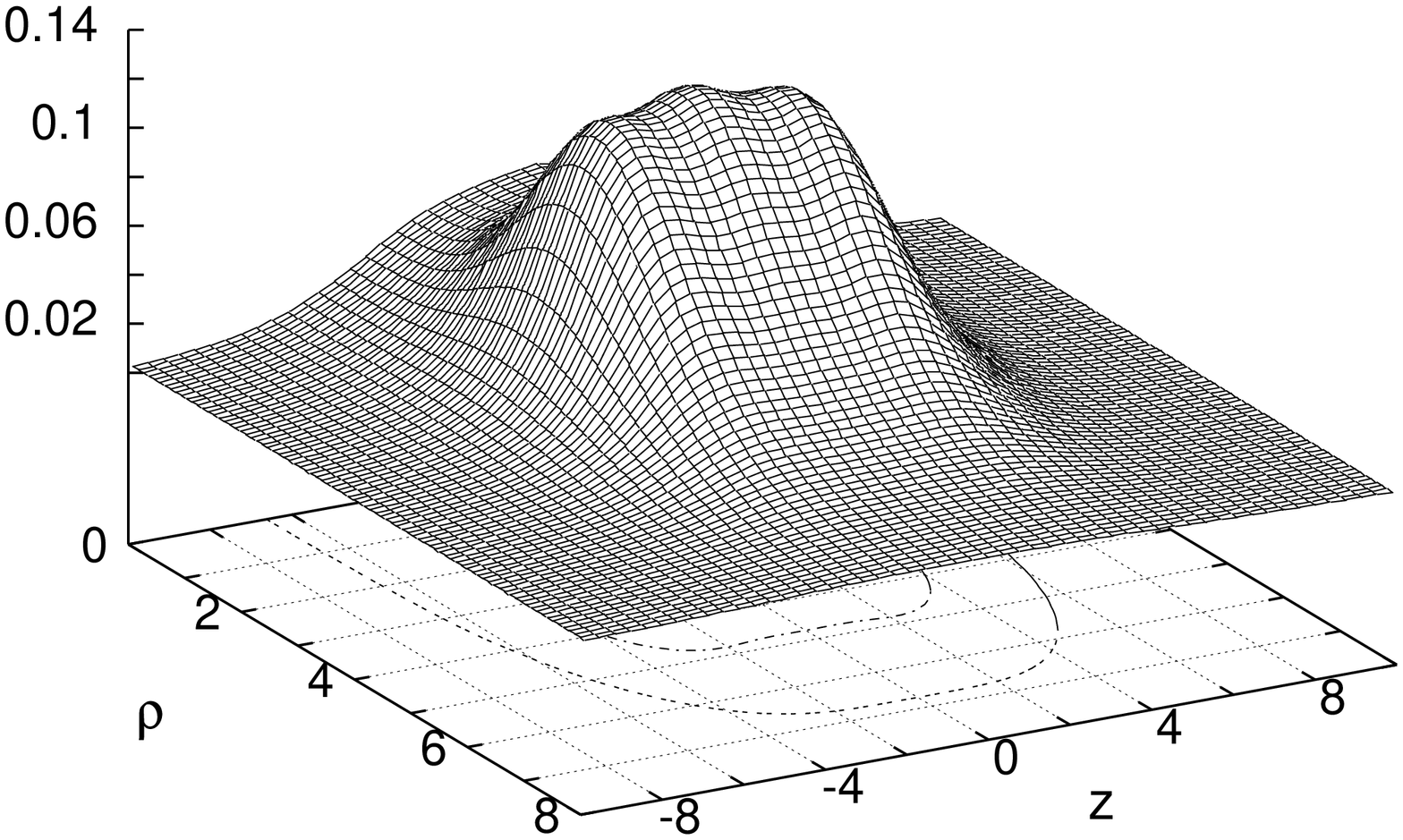}
  \includegraphics[height=.30\textheight, angle =-90]{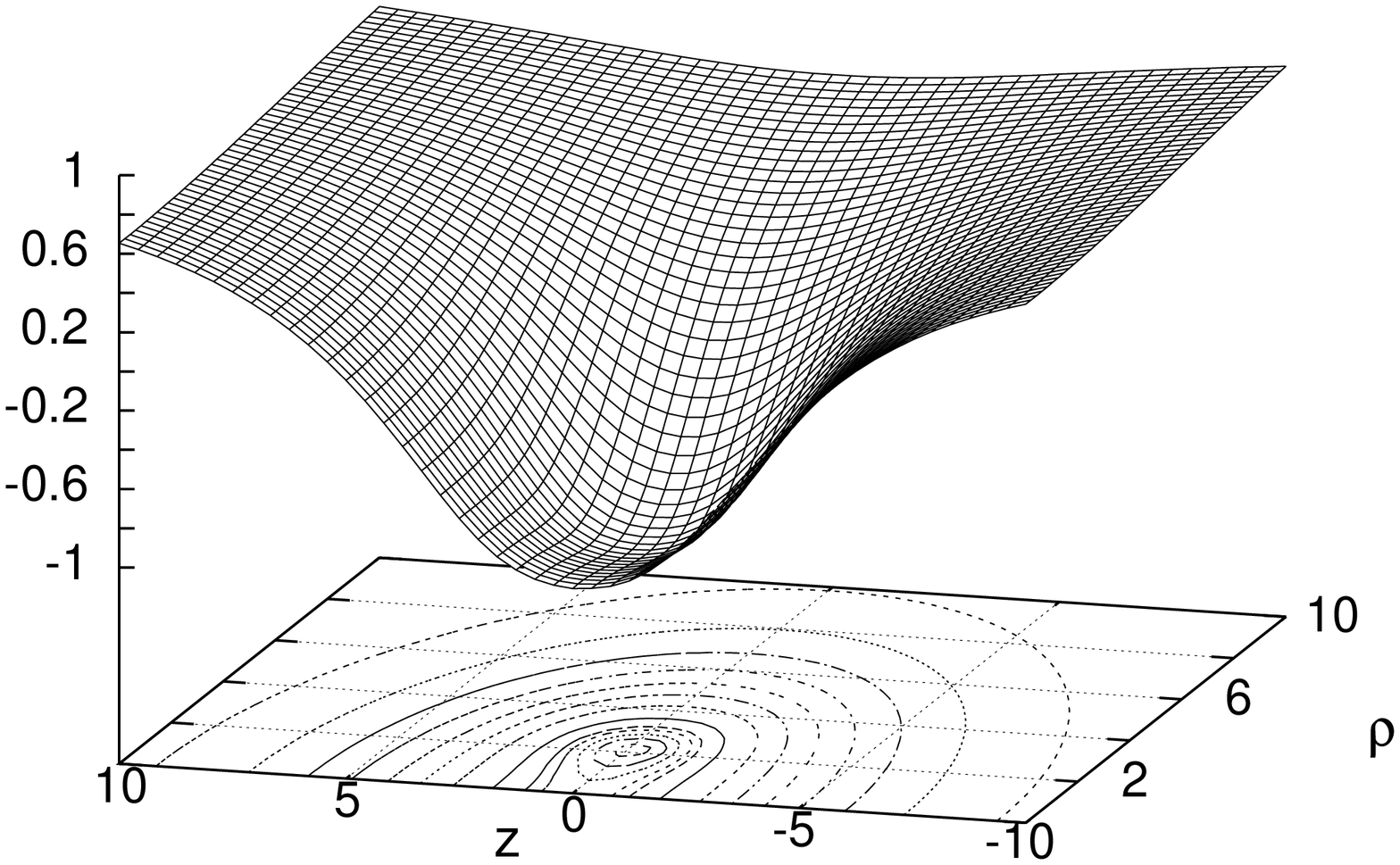}
  \includegraphics[height=.30\textheight, angle =-90]{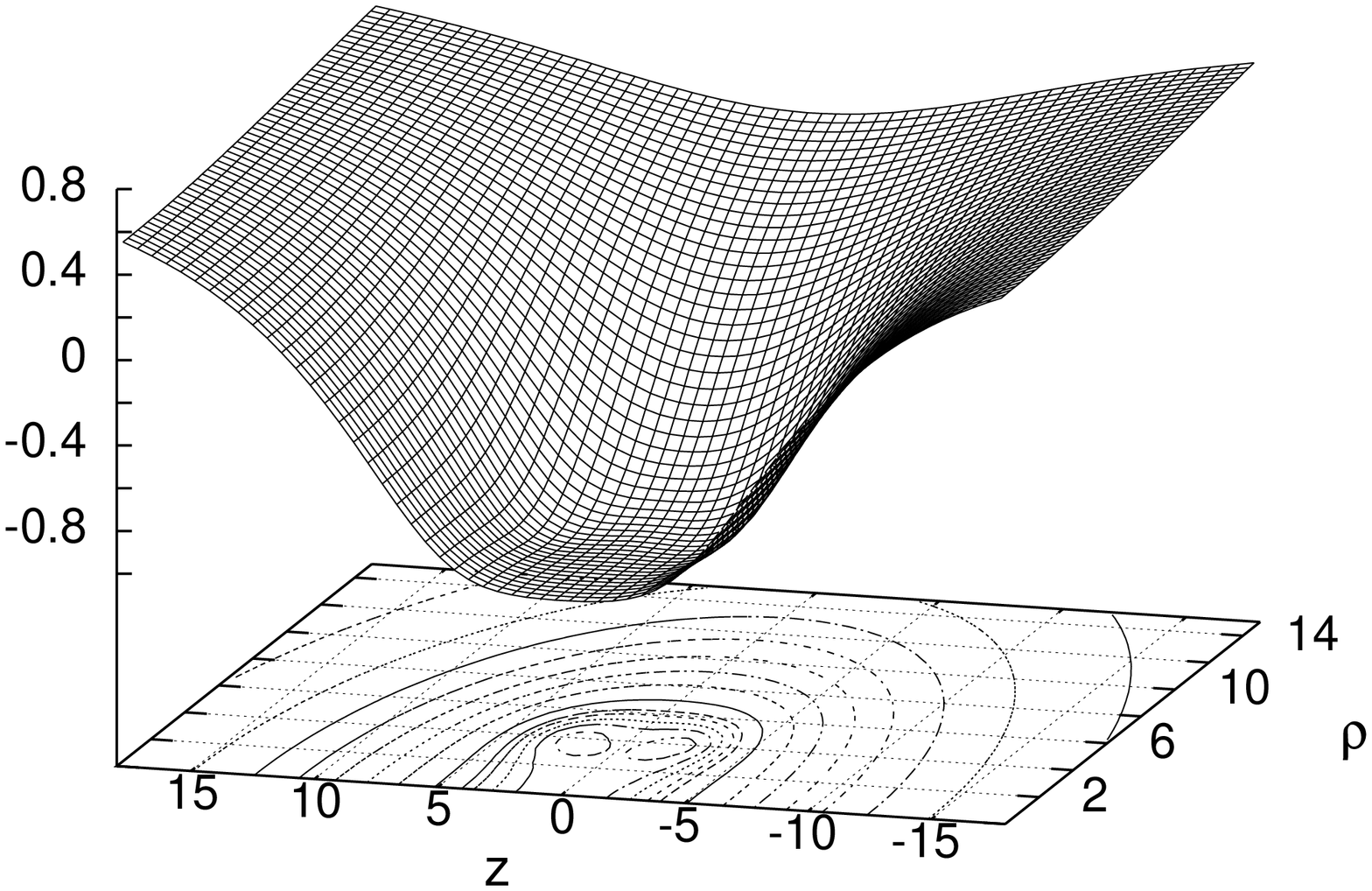}
\end{center}
\vspace{-0.2cm}
\caption{The action density plotted for the $n=3$, $m=2$ (upper left) 
and $n=3$, $m=3$ (upper right) caloron chain solutions.  
The Polyakov loop is plotted for the $n=3$, $m=2$ (bottom left) caloron
and compared with the $n=4$, $m=4$ (bottom right) solution.}
\label{fig:figure3}
\end{figure}

The same general behavior is observed for the other solutions.
Generally, the winding number $n$ is related to the (integer) 
topological charge of each individual pseudoparticle. Increasing $n$ 
to $n > 1$ deforms the local maxima of the action density into tori 
around the symmetry axis with a nonvanishing (and increasing with $n$) 
radius. 
For example, a single ring is formed for the configuration with 
$n=3$ and $m=2$ (see the upper plots of Fig. \ref{fig:figure3}).
Similarly, the local Polyakov loop now takes on the value  
${\cal P}({\vec x})=-1$ on circles in the core of the torus around the 
symmetry axis with radius $\rho_n$ (see the bottom plots 
of Fig. \ref{fig:figure3}) where 
the action density has a toroidal maximum. 
For a $Q=3$ instanton-antiinstanton 
configuration with $n=m=3$ we found three maxima of the action density 
located in three $x-y-$planes corresponding to three tori, one sitting 
at $z=0$ and the other two sitting symmetrically at $z = \pm \Delta z$ 
(see the right upper plot of Fig. \ref{fig:figure3}). 
The local holonomy precisely mimicks this behavior.

%%%%%%%%%%%%%%%%%%%%%%%%%%%%%%%%%%%%%%%%%%%%%%%%%%%%%%%%%%%%%%%%%%%%%
%%%%%%%%%%%%%%%%%%%%%%%%%%%%%%%%%%%%%%%%%%%%%%%%%%%%%%%%%%%%%%%%%%%%%

\section{Conclusions}
We have constructed axially symmetric caloron solutions for the 
four-dimensional Euclidean $SU(2)$ YM theory by numerical solution 
of the second order Yang-Mills equations with trivial asymptotic 
holonomy. Similarly to the axially symmetric monopole-antimonopole 
and vortex ring solutions of the YMH theory, the calorons are 
labeled by two winding numbers, $n$ and $m$. They are consisting of 
pseudoparticles of topological charge $\pm n$ building up a total 
topological charge $Q = \frac{n}{2} \left[1-(-1)^m\right]$. 
The action density of the configuration has a non-trivial shape.  
The position of its maxima allow us to identify a pointlike location 
or a toroidal shape of each individual constituent, depending on the
value of $n$. We have studied 
here also the landscape of the holonomy or Polyakov loop. In all cases
the loci of ${\cal P}({\bf r_0})=-1$ coincide with the maxima of the action 
density.
\vspace{-0.3cm}

\section*{Acknowledgements}
We are grateful to Pierre van Baal, Falk Bruckmann, Michael 
M\"uller-Preussker, Eugen Radu and Tigran Tchrakian for useful 
discussions and comments. This work was carried out in the framework 
of the Research Frontiers Programme of the Science Foundation of 
Ireland, project RFP07-330PHY. Ya.~S. asknowledges the travel
support by the Alexander von Humboldt Foundation.
E.-M. I. was supported by DFG through the Forschergruppe FOR 465 
(Mu 932/2). He is grateful to the Karl-Franzens-Universit\"at Graz 
for the guest position he holds while this talk is written up.

\end{document}